\documentclass[aps,pra,nofootinbib,superscriptaddress,showpacs,reprint,floatfix]{revtex4-1}
\usepackage{graphicx}
\usepackage{amsmath}
\usepackage{subfigure}
\usepackage{bm}
\usepackage{verbatim}
\usepackage[colorlinks=true, pdfstartview=FitV,urlcolor=blue,linkcolor=orange,citecolor=cyan]{hyperref}
\usepackage{color}
\usepackage{xcolor}
\usepackage{hyperref}
\usepackage{amssymb}
\usepackage{bbm}
\usepackage{dsfont}
\usepackage{siunitx}
\usepackage{txfonts}
\usepackage{physics}
\usepackage{soul}

\newcommand{\bk}{\bm{k}}

\newcommand{\average}[1]{\langle#1\rangle}

\newcommand{\TwoDMatrix}[4]{\begin{pmatrix} #1 & #2 \\ #3 & #4 \end{pmatrix}}
\newcommand{\TBG}{T_\text{BG}}
\newcommand{\vph}{v_\text{ph}}
\newcommand{\zinf}{z_{\infty}}
\newcommand{\phik}{\phi_{\bk}}
\newcommand{\tdh}{\tilde{h}}
\newcommand{\Moire}{{moir\'{e}}}
\usepackage{mathtools}

\newcommand{\BGBGBG}{Bloch-Gr\"{u}neisen}

\DeclarePairedDelimiterX\inner[2]{\langle}{\rangle}{#1|#2}

\allowdisplaybreaks[4]

\begin{document}
\title{Phonon scattering induced carrier resistivity in twisted double bilayer graphene}

\author{Xiao Li}
\email{xiao.li@cityu.edu.hk}
\affiliation{Department of Physics, City University of Hong Kong, Kowloon, Hong Kong SAR}
\affiliation{Condensed Matter Theory Center and Joint Quantum Institute, University of Maryland, College Park, MD 20742, USA}
\author{Fengcheng Wu}
\affiliation{Condensed Matter Theory Center and Joint Quantum Institute, University of Maryland, College Park, MD 20742, USA}
\author{S. Das Sarma}
\affiliation{Condensed Matter Theory Center and Joint Quantum Institute, University of Maryland, College Park, MD 20742, USA}

\date{\today}

\begin{abstract}
In this work we carry out a theoretical study of the phonon-induced resistivity in twisted double bilayer graphene (TDBG), in which two Bernal-stacked bilayer graphene devices are rotated relative to each other by a small angle $\theta$.
We show that at small twist angles ($\theta\sim 1^\circ$) the effective mass of the TDBG system is greatly enhanced, leading to a drastically increased phonon-induced resistivity in the high-temperature limit where phonon scattering leads to a linearly increasing resistivity with increasing temperature.
We also discuss possible implications of our theory on superconductivity in such a system, and provide an order of magnitude estimation of the superconducting transition temperature.
\end{abstract}

\maketitle

\section{Introduction}
Recent experimental discoveries of correlated insulator and superconductivity in twisted bilayer graphene (TBG)~\cite{Cao2018,Cao2018a,Yankowitz2019,Sharpe2019,Lu2019} have attracted great interest in the community.
The fact that the electronic band structure of TBG can become almost flat near the magic angle~\cite{Bistritzer2011} strongly enhances the effect of interactions, making it possible to study novel quantum phases that are otherwise difficult to realize experimentally.
Furthermore, the apparent similarities between the phase diagram of TBG and cuprate high-temperature superconductors~\cite{Lee2006} suggest that the study of electron correlations in TBG may provide useful hints for our understanding of the electronic properties in cuprates.

The experimental observation of novel quantum phases in TBG has since stimulated the investigation of other van der Waals heterostructures using the twist angle degree of freedom, including, e.g., the trilayer graphene/h-BN {\Moire} superlattice~\cite{Chen2019,Chen2019a}.
One of the motivations of such studies is to go beyond certain limitations of TBG.
For example, although the twist angle offers an unprecedented tuning knob to modify the electron band structure in TBG, it still cannot be changed continuously.
To date, properties of TBG are mostly modified by fabricating new devices with different twist angles or by applying hydrostatic pressure~\cite{Carr2018, Yankowitz2019}.
It will thus be advantageous to find a way to modify the band structure of a van der Waals heterostructure continuously near quantum critical points, which will enable a more detailed experimental characterization of the electron correlation effects.

Twisted double bilayer graphene (TDBG) has emerged as a promising platform in this respect, because the band structure of a single Bernal-stacked bilayer graphene~\cite{Neto2009,McCann2013} can be tuned continuously by an external perpendicular electric field.
Consequently, one can expect to adjust the band structure of TDBG continuously by an external electric field.
Such a tunability is highly desirable, especially near certain quantum critical points.
As a result, TDBG has attracted much attention and rapid experimental~\cite{Shen2019,Cao2020,Liu2019,Burg2019} and theoretical~\cite{Zhang2019,Lee2019} progress has been made.
In particular, the application of an external electric field has indeed given rise to a very rich TDBG phase diagram, including signatures of correlated insulator states as well as superconductivity and possibly ferromagnetism in some cases.

The interesting TDBG physics for small twist angles arises {from the same moire flatband physics dominating the extensively studied TBG phenomena near the magic twist angle}. 
Basically, the moire potential for small twist angle strongly flattens the relevant graphene bands, leading to very small band velocities (or very large carrier effective masses), which lead to a great enhancement of all interaction phenomena since typically interaction physics is proportional to the carrier effective mass.  
In the current work, we use a suitable continuum model TDBG band structure to estimate band flattening effects.

In this work, however, we take a different perspective and study the resistivity in TDBG in the high-temperature limit, when phonon scattering will be the dominant mechanism for resistivity.
Thus, instead of focusing on the $T=0$ ground state phase diagram, we investigate the ohmic transport properties of the finite temperature effective metallic TDBG phase above the applicable critical temperatures (or the ground state energy gaps) of the symmetry-broken states where TDBG behaves as a metal. Considering rather clean systems, and focusing specifically on the temperature dependence of carrier resistivity, we neglect effects of disorder, impurities, and defects since the main temperature dependence of metallic resistivity arises from phonon scattering effects. We also ignore all electron-electron interaction effects, and only take into account resistive scattering by acoustic phonon scattering.

Specifically, the scattering of electrons by acoustic phonons can be generally divided into two regimes: a low-temperature regime ($T<\TBG$) and a high-temperature one ($T>\TBG$). The characteristic temperature $\TBG$ is known as the {\BGBGBG}  (BG) temperature, given by $k_B\TBG = 2\hbar \vph k_F$~\cite{Hwang2008,Min2011}, where $k_B$ is the Boltzmann constant, $\vph$ is the phonon velocity and $k_F$ is the Fermi wave vector.
(We note that for regular metals where $\TBG$ is very high, or in any situation where $\TBG>T_D$ with $T_D$ being the Debye temperature, the characteristic temperature defining the low and high temperature phonon scattering regimes is $T_D$ and not $\TBG$.)
In the high-temperature regime, the electron resistivity will scale as a linear function of temperature $T$, giving rise to $\rho\sim T$, which has been well understood in the context of graphene devices~\cite{Neto2009,Peres2010,Sarma2011}.
We are interested in this regime because similar to TBG, a wide range of linear-in-$T$ resistivity has been observed in this temperature range in TDBG~\cite{Shen2019,Cao2020,Liu2019}.
In the context of TBG, such a behavior is often attributed to the putative `strange-metal' phase~\cite{Cao2019a,Polshyn2019}, although it can be compatible with a phonon-scattering mechanism~\cite{Wu2019}, 
{albeit} with greatly enhanced phonon scattering induced carrier resistivity.
In this work, we will theoretically study the phonon-induced resistivity in TDBG in the high-temperature regime, and analyze its compatibility with the experimental observations.
In particular, we want to understand whether electron-phonon scattering in TDBG can be a contributing factor for the linear-in-$T$ resistivity seen in recent experiments.
Our work can be thought of as the TDBG generalization of Ref.~\cite{Wu2019} or as the small twist angle double-bilayer generalization of Ref.~\cite{Min2011}.
The goal is to theoretically obtain the acoustic phonon scattering induced carrier resistivity of TDBG as a function of temperature, twist angle, and carrier density. 

The structure of the paper is the following.
In Section~\ref{Section:Model} we set up a continuum model for TDBG and demonstrate that its low-energy bands become almost flat (i.e. very large effective mass or equivalently very small effective velocity) at small twist angles ($\theta\sim 1^\circ$).
In Section~\ref{Section:Resistivity} we explain the theoretical framework we use to evaluate phonon-induced resistivity in TDBG and present our numerical results.
In Section~\ref{Section:Discussions} we provide some additional discussions. In particular, we will comment on the possible implications of our theory on superconductivity in TDBG, and provide a rough estimate of the superconducting transition temperature $T_c$ arising from the enhanced electron-phonon coupling.
Finally, in Section~\ref{Section:Conclusions} we provide a brief summary of our results.

\section{Continuum description of a twisted double bilayer graphene \label{Section:Model}}
\begin{figure}[t]
\includegraphics[width = 8 cm]{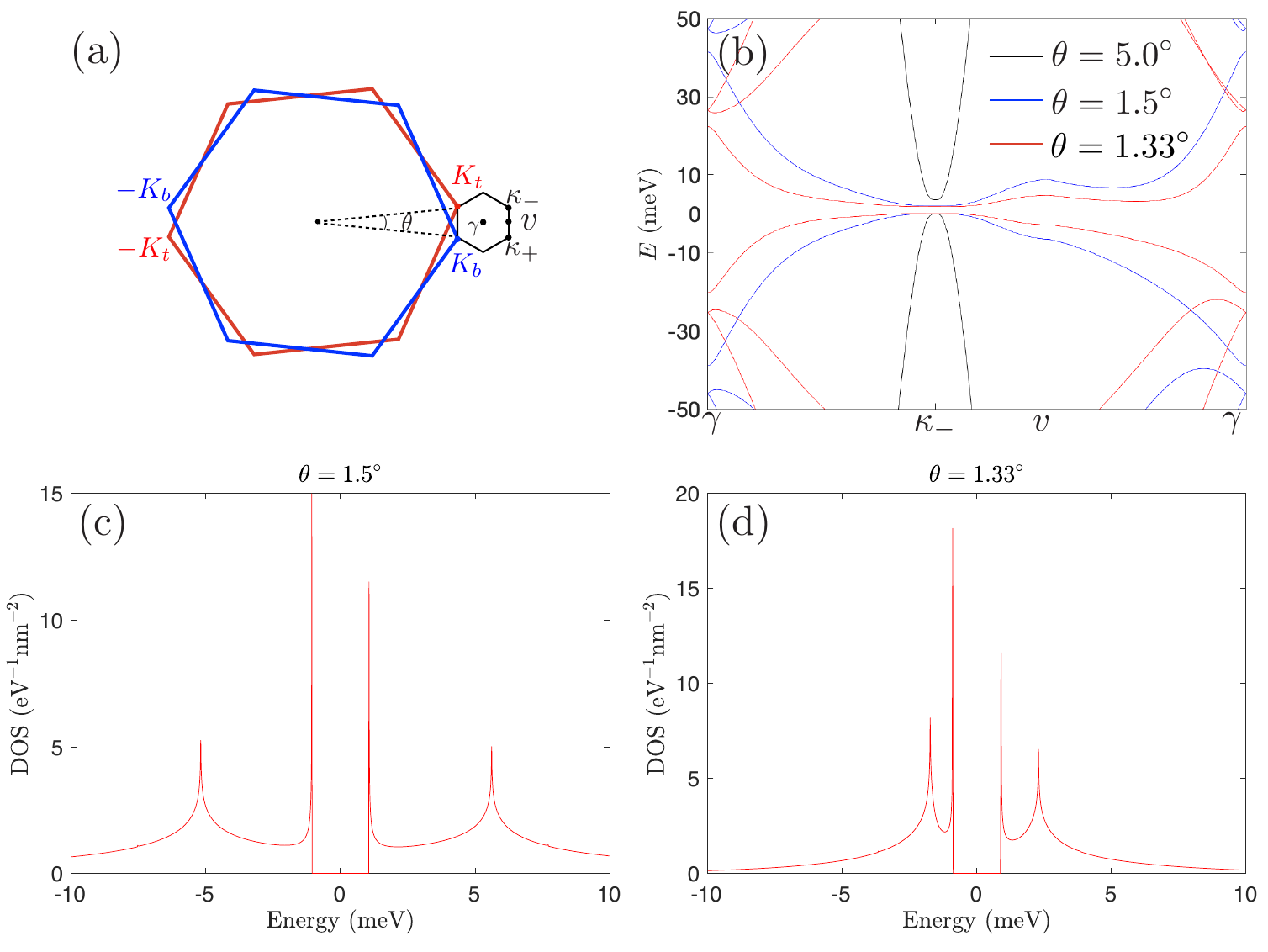}
\caption{\label{Fig:Bandstructure}
(a) Illustration of the {\Moire} Brillouin zone in twisted double bilayer graphene. Here, the red and blue hexagon denotes the Brillouin zone of the top and bottom bilayer graphene, respectively, while the black hexagon represents the {\Moire} Brillouin zone.
(b) Band structure of twisted double bilayer graphene at various different twist angles. Here we choose $\gamma_1 = \SI{380}{meV}$, $w = \SI{118}{meV}$.
(c)-(d) Density of states per spin per valley in twisted double bilayer graphene at a twist angle of $\theta = 1.5^\circ$ and $\theta = 1.33^\circ$, respectively.
}
\end{figure}

We start by introducing the continuum model of a TDBG.
We consider two Bernal-stacked bilayer graphene (BLG) rotated relative to each other by a small angle $\theta$, as shown in Fig.~\ref{Fig:Bandstructure}(a).
In particular, we adopt the convention that the top BLG will be rotated by an angle of $\theta/2$, while the bottom one will be rotated by $-\theta/2$.
As a result, the continuum description of TDBG near valley $+K$ can be written as
\begin{align}
	\mathcal{H}_{+} = \TwoDMatrix{h_t(\bk)}{T(\bm{r})}{T^\dagger(\bm{r})}{h_b(\bk)}, \label{Eq:TheModel}
\end{align}
which is given in the basis of \{$A_1$, $B_1$, $A_2$, $B_2$, $A_3$, $B_3$, $A_4$, $B_4$\}.
Here $A$ and $B$ denote the two sublattices of BLG and indices $1$-$4$ denote the four atomic layers, with $1$-$2$ belonging to the top BLG and $3$-$4$ to the bottom one. 

In the continuum model Eq.~\eqref{Eq:TheModel}, $h_{t(b)}$ denotes the Hamiltonian for the isolated top (bottom) BLG, given by
\begin{align}
	h_{\lambda}(\bk) =
	\begin{pmatrix}
		0 & \hbar vk_{\lambda}^\ast e^{il_{\lambda}\theta/2} & 0 & 0  \\
		\hbar vk_{\lambda}e^{-il_{\lambda}\theta/2} & 0 & \gamma_1 & 0\\
		0 & \gamma_1 & 0 & \hbar vk_{\lambda}^\ast e^{il_{\lambda}\theta/2}\\
		0 & 0 & \hbar vk_{\lambda}e^{-il_{\lambda}\theta/2} & 0
	\end{pmatrix}. \label{Eq:BareBLG}
\end{align}
In the above equation $\lambda = t,b$ denotes the top and bottom BLG, and $l_{t(b)}=+1 (-1)$.
In addition, $k_{t(b)}\equiv k_x^{(t(b))} + ik_y^{(t(b))}$ denotes the (complex) in-plane momentum measured from the Brillouin zone corner of the top (bottom) BLG.
In addition, $v = \SI{1e6}{m/s}$ is the bare Dirac velocity of monolayer graphene, while $\gamma_1$ is the interlayer coupling energy of an isolated BLG.
Note that the value of $\gamma_1$ in the literature varies widely from $\SI{300}{meV}$ to $\SI{400}{meV}$~\cite{McCann2013}.
In this work we will take $\gamma_1 = \SI{380}{meV}$, but note that results do depend on the specific choice of the $\gamma_1$ band parameter.

In TDBG the {\Moire} potential arising from the twist angle between the two BLGs only induces direct coupling between atomic layers $2$ and $3$.
As a result, the {\Moire} potential term in the continuum model Eq.~\eqref{Eq:TheModel} can be written as
\begin{align}
	T(\bm{r}) =
	\begin{pmatrix}
		0 & 0 \\ \mathbbm{t}(\bm{r}) & 0
	\end{pmatrix},
\end{align}
where $\mathbbm{t}(\bm{r}) = w \displaystyle\sum_{j=1}^3 T_j e^{i\bm{Q}_j\cdot \bm{r}}$.
Here $w\simeq\SI{118}{meV}$~\cite{Wu2018} and
\begin{align}
	T_j = \sigma_0 + \cos(2\pi j/3)\sigma_x + \sin(2\pi j/3)\sigma_y, \; (j = 1, 2, 3),
\end{align}
where $\sigma_i$ are the Pauli matrices.
The three vectors $\bm{Q}_j$ read as
\begin{align}
	\bm{Q}_1 = K_\theta \left(\dfrac{\sqrt{3}}{2}, \dfrac{1}{2}\right), \;
	\bm{Q}_2 = K_\theta \left(-\dfrac{\sqrt{3}}{2}, \dfrac{1}{2}\right), \;
	\bm{Q}_3 = K_\theta (0, -1), \notag
\end{align}
with $K_\theta = 4\pi/(3a_M)$. Here $a_M = a_0/[2\sin(\theta/2)]$ is the lattice constant of TDBG, and $a_0=\SI{2.46}{\AA}$ is the lattice constant of monolayer graphene.

\begin{figure}[b]
\includegraphics[width = 8cm]{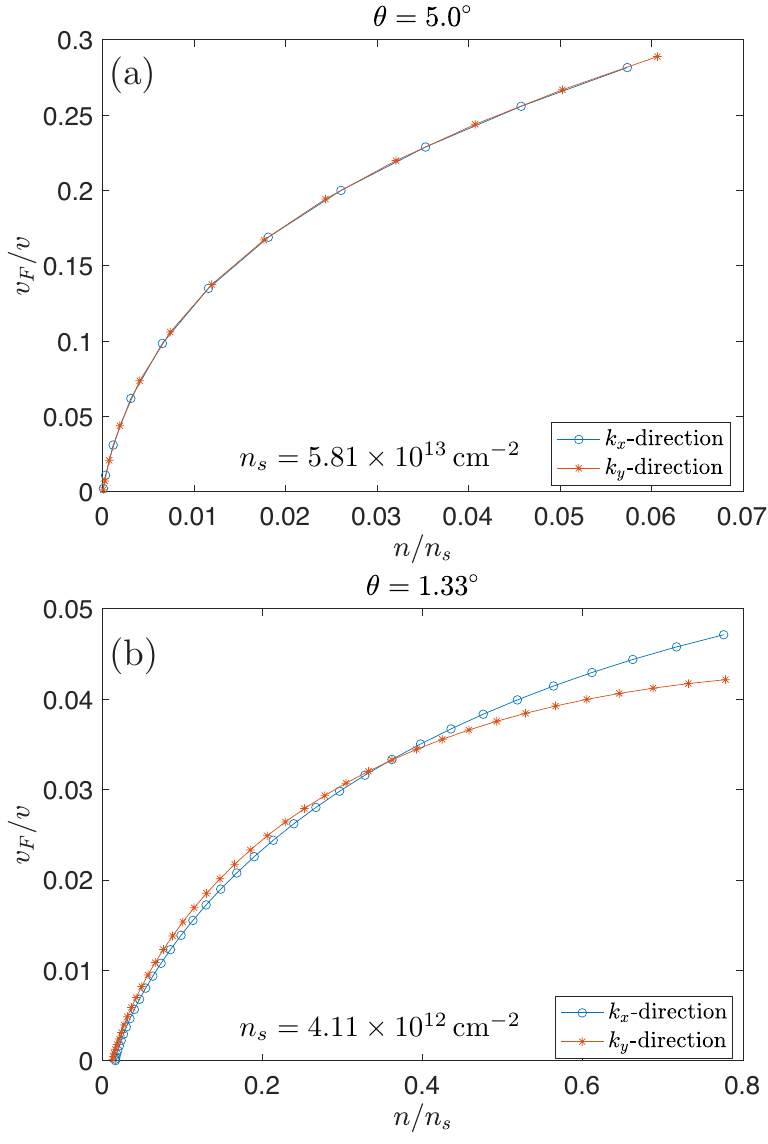}
\caption{\label{Fig:FermiVelocity} Fermi  velocity $v_F$ of TDBG for a twist angle of (a) $\theta=5.0^\circ$ and (b) $\theta=1.33^\circ$. The horizontal axes in the two panels are chosen so that the actual density range is similar. 
The red and blue lines represent the Fermi velocity along $k_x$ and $k_y$ directions, respectively. The set of parameters used in this figure is the same as Fig.~\ref{Fig:Bandstructure}. In particular, $v = \SI{1e6}{m/s}$ is the Fermi velocity of a pristine monolayer graphene. Moreover, the corresponding value of $n_s$ (see text) in each case is shown in the figure. }
\end{figure}

Within this continuum description, the band structure of TDBG can be obtained by diagonalizing a large matrix in the momentum space which connects each $\bk$ point in the first superlattice {\Moire} Brillouin zone (MBZ) to three other points $\bk + \bm{Q}_j$ ($j = 1,2,3$) in an adjacent MBZ.
The resulting band structures for TDBG at three different twist angles are shown in Fig.~\ref{Fig:Bandstructure}(b).
For a large twist angle ($\theta=5.0^\circ$), the band structure near the $\kappa_{-}$ point in the MBZ is close to that of pristine BLG, although a small band gap opens up due to the broken inversion symmetry.
In contrast, for much smaller twist angles ($\theta = 1.5^\circ$ and $\theta = 1.33^\circ$), the bands near the $\kappa_{-}$ point become very flat, giving rise to a large density of states (DOS) near the band bottom.
In Fig.~\ref{Fig:Bandstructure}(c)-(d) we show the DOS per spin per valley $\nu(\varepsilon_F)$ in TDBG, which indeed becomes quite large for small twist angles ($\theta\sim1^\circ$).
In addition, the flattened bands also lead to a much reduced Fermi velocity, as shown in Fig.~\ref{Fig:FermiVelocity}.
This physics strongly enhances phonon scattering as we discuss later in the paper.
{Note that in this figure we have introduced the total electron density $n_s$ of a filled conduction band, taking into account both the spin and the valley degeneracies. 
Its explicit expression reads as $n_s=8/\qty(\sqrt{3}a_M^2)$~\cite{Burg2019}. 
In particular, we find $n_s=\SI{5.81e13}{cm^{-2}}$ when $\theta=5.0^\circ$, and $n_s=\SI{4.11e12}{cm^{-2}}$ when $\theta=1.33^\circ$. 
Both results in Fig.~\ref{Fig:FermiVelocity} are plotted against the respective $n_s$ in order to provide a context for the corresponding electron densities. 
} 

\subsection{Approximate zero-energy eigenstates}
In order to obtain the full band structure of TDBG one must numerically diagonalize a large matrix.
However, near the $\kappa_{\pm}$ points in the MBZ it is possible to obtain an approximate analytical expression for the two lowest-energy eigenstates.
Such a method was first developed in Ref.~\cite{Bistritzer2011} to obtain an approximate two-band model for TBG, and was used to obtain approximate lowest-energy eigenstates in TDBG in Ref.~\cite{Zhang2019}.

Specifically, one can truncate the Hamiltonian $\mathcal{H}_+$ in Eq.~\eqref{Eq:TheModel} by retaining only four momentum points $\bk_t$, and $\bk_b^{(j)} \equiv \bk_t + \bm{Q}_j$ ($j = 1, 2, 3$), and obtain a $8$-by-$8$ Hamiltonian $H(k)$ as follows: 
\begin{align}
	H(k) =
	\begin{pmatrix}
		\tdh_0(k) & T_1(k) & T_2(k) & T_3(k) \\
		T_1^{\dagger}(k) & \tdh_1(k) & 0 & 0 \\
		T_2^{\dagger}(k) & 0 & \tdh_2(k) & 0 \\
		T_3^{\dagger}(k) & 0 & 0 & \tdh_3(k) \\
	\end{pmatrix}, \label{Eq:8-bandModel}
\end{align}
where~\cite{Zhang2019}
\begin{align}
	\tdh_0(k) &=
	\begin{pmatrix}
		0 & -v^2(k_\theta^\ast)^2/\gamma_1 \\
		-v^2(k_\theta)^2/\gamma_1 & 0
	\end{pmatrix}, \notag\\
	\tdh_j(k) &=
	\begin{pmatrix}
		0 & -v^2\left[(k+Q_j)_{-\theta}^\ast\right]^2/\gamma_1 \\
		-v^2\left[(k+Q_j)_{-\theta}\right]^2/\gamma_1 & 0
	\end{pmatrix}, \notag\\
	T_j(k) &=
	\begin{pmatrix}
		-t_Mvk_\theta^\ast/\gamma_1 & t_Mv^2\lambda_j^\ast k_\theta^\ast(k+Q_j)_{-\theta}^\ast/\gamma_1^2\\
		t_m\lambda_j & -t_Mv(k+Q_j)_{-\theta}^\ast/\gamma_1
	\end{pmatrix}.
\end{align}
In the above results, we have defined $\lambda_j = e^{2\pi ij/3}$ ($j = 1, 2, 3$).
In addition, we have introduced a short-hand notation that $k_\theta \equiv (k_x + ik_y)e^{-i\theta/2}$.

One can verify that in the $k\to0$ limit the two zero-energy eigenstates of $H(k)$ in Eq.~\eqref{Eq:8-bandModel} can be approximately written as 
\begin{align}
	\ket{\Psi^{(\alpha)}} = S_\alpha
	\begin{pmatrix}
		\psi_0^{(\alpha)} \\ \psi_{1}^{(\alpha)}\\ \psi_{2}^{(\alpha)} \\ \psi_3^{(\alpha)}
	\end{pmatrix},
	\quad \alpha = A, B, \label{Eq:ZeroEnergyState}
\end{align}
where $S_\alpha$ is the normalization factor, $\psi_0^{(\alpha)}$ is a two-component spinor, and
\begin{align}
	\psi_j^{(\alpha)} = \dfrac{t_M}{vK_\theta^4}\TwoDMatrix{0}{K_\theta^2(Q_j)_\theta}{0}{\dfrac{\gamma_1}{v}e^{ij\phi}\left[(Q_j)_\theta^2\right]^\ast} \psi_0^{(\alpha)} \equiv M_j \psi_0^{(\alpha)}, \quad j = 1, 2, 3. \notag
\end{align}
The wave function normalization can be determined from the following condition,
\begin{align}
	1 &= \inner{\Psi^{(\alpha)}}{\Psi^{(\alpha)}}
	= \lvert S_\alpha\rvert^2 \left(\psi_0^{(\alpha)}\right)^\dagger \left[1 + \sum_{j=1}^{3} M^\dagger_j M_j\right]\psi_0^{(\alpha)} \notag\\
	&= \lvert S_\alpha\rvert^2 \left(\psi_0^{(\alpha)}\right)^\dagger \TwoDMatrix{1}{0}{0}{1+3\Delta} \psi_0^{(\alpha)},
\end{align}
where $\Delta = \dfrac{\gamma_1^2t_M^2}{v^4K_\theta^4}\left(1+\dfrac{v^2K_\theta^2}{\gamma_1^2}\right)$.
Now if we adopt the natural choice of $\psi_0^{(A)} = \begin{pmatrix} 1\\0\end{pmatrix}$, and $\psi_0^{(B)} = \begin{pmatrix} 0\\ 1 \end{pmatrix}$, we find that
\begin{align}
	S_A = 1, \quad S_B = \dfrac{1}{\sqrt{1+3\Delta}}. \label{Eq:Normalization}
\end{align}
This approximate analytical expression for the lowest-energy states in TDBG will be helpful for our analysis of phonon-induced resistivity in the next section.

\section{Phonon-induced resistivity \label{Section:Resistivity}}
In the previous section we have shown that for small twist angles ($\theta\sim1^\circ$), the Fermi velocity of TDBG near the MBZ corners can become quite small.
As we will show in this section, such a  substantial reduction in Fermi velocity can give rise to a much enhanced phonon-induced resistivity in the high-temperature ($T\gg\TBG$) limit as happens also for TBG at small twist angles~\cite{Wu2019}.
In particular, in such a limit the resistivity scales linearly with temperature, $\rho\approx CT$, and our goal is to estimate the coefficient $C$ and explain how it increases substantially at small twist angles.
To verify the validity of our theory, we also numerically evaluate the resistivity in the full temperature range (i.e. $T\gg\TBG$ as well as $T\ll\TBG$), and estimate the crossover temperature above which this enhanced linear-in-$T$ resistivity regime applies.

\subsection{Resistivity from Boltzmann transport theory}
To begin with, we recall that in the Boltzmann transport theory the energy-averaged scattering time $\tau$ in monolayer graphene in the limit of $k_BT\ll \varepsilon_F$ is given by~\cite{Min2011}
\begin{align}
	\average{\tau}^{-1} = \dfrac{2\pi}{\hbar}\nu_0\lvert W(k_F)\rvert^2 I, \label{Eq:ScatteringTime}
\end{align}
where $\nu_0$ is the DOS per spin and valley at the Fermi energy, and $\lvert W(k_F)\rvert^2 = D^2\hbar k_F/(2\rho_m \vph)$ is the squared matrix element for acoustic phonon scattering.
Here $\rho_m=\SI{7.6e-8}{g/cm^2}$ is the mass density of a single graphene sheet, $\vph=\SI{2.6e6}{cm/s}$ is the velocity of longitudinal acoustic (LA) phonon in monolayer graphene, $D=\SI{25}{eV}$ is the acoustic phonon deformation potential~\cite{Min2011}, and $v_F$ is the Fermi velocity.
The integral $I$ has the following form,
\begin{align}
	I = \int \dfrac{\phi}{2\pi} \dfrac{F(q)(1-\cos\phi)}{\epsilon^2(q)} \dfrac{2q}{k_F}\beta \hbar\omega_q N_q(N_q+1), \label{Eq:IntegralI}
\end{align}
where $q = 2k_F \sin(\phi/2)$ is the magnitude of the acoustic phonon wave vector, $\beta = 1/(k_B T)$, and $F(q)$ is the chiral factor defined as the square of the wave function overlap between incoming and scattered electrons.
In addition, $N_q = (e^{\beta\hbar\omega_q}-1)^{-1}$ is the phonon occupation number, with $\omega_q = \vph q$ being the  frequency of the acoustic phonon.
Finally, $\epsilon(q)$ is the dielectric function, which takes into account the screening effect at wave vector $q$. In this work we will only consider the unscreened limit, so we will take $\epsilon(q) = 1$.
The reason for neglecting screening, which is easy to include, is that there is no experimental evidence that the electron-acoustic phonon resistive scattering gets screened in graphene as a direct comparison between theory~\cite{Hwang2008} and experiment~\cite{Efetov2010} supports the unscreened approximation.
We thus do not believe that screening plays any role in TDBG (or TBG) phonon scattering.

{
Before we proceed, we note that in this work we have left out two important aspects of the phonons in TDBG.
First, we did not account for the possible modification of electron-phonon coupling when the twist angle is varied~\cite{Choi2018}, which can be strong near small twist angles. 
Second, we only consider the contributions to the resistivity from longitudinal acoustic (LA) phonons, as it has been shown to be the main contribution in monolayer graphene at low temperatures~\cite{Hwang2008}.
However, because TDBG has lower symmetry than monolayer graphene does, it is possible that the out-of-plane phonon modes~\cite{Park2014}, i.e., the transverse acoustic (TA) and the parabolic ZA mode, may also contribute to the resistivity in TDBG.
A more comprehensive modeling for phonons in TDBG should account for these effects.
However, we do not attempt to present such a model in this work because currently there is considerable sample to sample variations in TDBG, which prevents the construction of a phonon model on a more quantitative level.
We hence leave such work for the future.
}

We now apply the above formalism to the case of TDBG, and obtain the electron resistivity as $\rho = \sigma^{-1}$, where $\sigma$ is the electron conductivity, given by
\begin{align}
  \sigma = g_s g_v e^2\nu(\varepsilon_F)\dfrac{v_F^2}{2}\average{\tau}. \label{Eq:Conductivity}
\end{align}
In the above equation $g_s = 2$ and $g_v = 2$ are the degeneracies due to electron spin and valley degrees of freedom, respectively, while $\nu(\varepsilon_F)$ is the DOS per spin per valley in TDBG shown in Fig.~\ref{Fig:Bandstructure}.
It is worth noting that when evaluating the scattering time $\average{\tau}$ in TDBG using Eq.~\eqref{Eq:ScatteringTime}, we should replace the DOS $\nu_0$ there by $\nu(\varepsilon_F)/2$, for the following reasons.
In this work we only consider electron densities below the van Hove singularity in TDBG, in which case the topology of the Fermi surface consists of two disconnected mini-valleys ($\kappa_{\pm}$) in the MBZ near the $+K$ valley in the original Brillouin zone of BLG.
As a result, the scattering matrix element $W(k_F)$ in Eq.~\eqref{Eq:ScatteringTime} is only appreciable for electrons within the same mini-valley.
The above observation leads us to conclude that in the low-density regime we are interested in, only half of electrons at the Fermi surface contribute to the scattering time.
Consequently, we need to substitute $\nu_0$ by $\nu(\varepsilon_F)/2$ in Eq.~\eqref{Eq:ScatteringTime}~\footnote{Note that we should still use the full DOS $\nu(\varepsilon_F)$ in Eq.~\eqref{Eq:Conductivity}, because all electrons at the Fermi surface contribute to the conductivity. }.
Putting everything together, we finally obtain the following expression for the resistivity in TDBG:
\begin{align}
 \rho =\dfrac{1}{2g_sg_v}\left(\dfrac{h}{e^2}\right)\left(\dfrac{D^2k_F I}{\hbar\rho_m\vph v_F^2}\right). \label{Eq:Resistivity}
\end{align}

In order to calculate the resistivity in TDBG using the above equation, we need to evaluate the integral $I$, whose explicit form is given in Eq.~\eqref{Eq:IntegralI}.
It can be simplified by setting $x = q/(2k_F) = \sin(\phi/2)$, which yields
\begin{align}
  I = \dfrac{16}{\pi}\int_{0}^{1} dx \dfrac{F(2k_Fx)}{\sqrt{1-x^2}} \dfrac{z_\text{BG} x^4 e^{z_\text{BG} x}}{(e^{z_\text{BG} x}-1)^2},
\end{align}
where $z_\text{BG} = \TBG/T$.
In the high-temperature limit ($T\gg\TBG$) we are interested in, we find that $I\approx z_{\infty}/z_\text{BG}$, where
\begin{align}
	\zinf = \dfrac{16}{\pi}\int_0^{1} dx \dfrac{x^2F(2k_Fx)}{\sqrt{1-x^2}},
\end{align}
and therefore the resistivity in Eq.~\eqref{Eq:Resistivity} becomes $\rho \approx CT$, where the coefficient $C$ is given by

\begin{align}
	C = \dfrac{\pi D^2k_Bz_{\infty}}{2g_s g_v e^2\hbar\rho_m\vph^2v_F^2}. \label{Eq:Coefficient_C}
\end{align}
Therefore, phonon-induced electron resistivity becomes linear in $T$ in the high-temperature ($T\gg\TBG$) limit, a regime we focus on in this work.
In addition, from the above result one can see that the quantity $z_\infty$ is a key quantity in this calculation, which depends solely on the chiral form factor $F(q)$ [or equivalently, $F(\phi)$]. Thus, we will discuss this quantity first.

\subsection{The chiral form factor}
We will use three different approximations to evaluate the chiral form factor $F(\phi)$ and hence $\zinf$ for low-energy conduction-band states in TDBG.
Specifically, we will use the two-band and four-band description for a pristine BLG, as well as a low-energy two-band description for TDBG.
We will see that they capture different aspects of the band structure.
A more accurate estimate of $F(\phi)$ necessitates a full numerical evaluation, which we leave for future studies.
We comment that, given the simplified nature of our TDBG band structure model, it is unclear that a full numerical calculation of the form factor is warranted.

\subsubsection{Two-band model for bilayer graphene}
We start with the simplest case, where a pristine BLG is described by a two-band model, given by
\begin{align}
	H_\text{BLG-2band} = -\dfrac{\hbar^2v^2}{\gamma_1}\TwoDMatrix{0}{(k_x+ik_y)^2}{(k_x-ik_y)^2}{0}.
\end{align}
As a result, the conduction band eigenstates are given by
\begin{align}
	\ket{\psi_{+}(\phik)} = \dfrac{1}{\sqrt{2}}\begin{pmatrix}e^{-2i\phik}\\1\end{pmatrix},
\end{align}
and the chiral form factor is given by
\begin{align}
	F(\phi) = \lvert\inner{\psi_{+}(\phik+\phi)}{\psi_{+}(\phik)}\rvert^2. \label{Eq:2BandFormFactor}
\end{align}
In order to evaluate $\zinf$, we note that $x = \sin(\phi/2)$, and thus $F(\phi) = (1-2x^2)^2$.
It follows that
\begin{align}
	\zinf^\text{(BLG-2band)} = \dfrac{16}{\pi}\int_0^1 dx \dfrac{x^2}{\sqrt{1-x^2}}(1-2x^2)^2 = 2. \label{Eq:zinf-BLG-2band}
\end{align}
As a result, within the two-band model of BLG, $\zinf$ is a constant, independent of either electron density or twist angle.

\subsubsection{Four-band model for bilayer graphene}
Next, we consider pristine BLG in the four-band description. The corresponding Hamiltonian is given by
\begin{align}
	H_\text{BLG-4band} =
	\begin{pmatrix}
		0 & \hbar vke^{-i\phik} & 0 & 0\\
		\hbar vke^{i\phik} & 0 & \gamma_1 & 0\\
		0 & \gamma_1 & 0 & \hbar vke^{-i\phik}\\
		0 & 0 & \hbar vke^{i\phik} & 0
	\end{pmatrix},
\end{align}
which is written in the \{$A_1$, $B_1$, $A_2$, $B_2$\} basis. We consider the lower conduction band of BLG, whose energy is
\begin{align}
	E_c(\bk) = \dfrac{1}{2}\left[\sqrt{4\hbar^2v^2k^2+\gamma_1^2} - \gamma_1\right],
\end{align}
and the corresponding wave function is given by
\begin{align}
	\psi&(\phik)^\dagger = \\
	&\begin{bmatrix}
		e^{2i\phik}\dfrac{-\sqrt{1+\eta}}{2}, & e^{i\phik}\dfrac{-\sqrt{1-\eta}}{2}, & e^{i\phik}\dfrac{\sqrt{1-\eta}}{2}, & \dfrac{\sqrt{1+\eta}}{2}
	\end{bmatrix}. \notag
\end{align}
In the above expression, $n$ is the electron density and $\eta = \left(1 + \frac{n}{n_0}\right)^{-1/2}$,
with $n_0 = k_0^2/\pi$ and $\hbar vk_0 = \gamma_1/2$.
From this wave function, we can obtain the chiral form factor $F(\phi)$ as follows,
\begin{align}
	F(\phi) = \dfrac{1}{4}\left[(1-\eta) + (1+\eta)\cos\phi\right]^2.
\end{align}
Note that in the low-density limit ($\eta\to1$) the above form factor reduces to $\cos^2\phi$, the result derived from the two-band model given in Eq.~\eqref{Eq:2BandFormFactor}, as expected.

The expression for $\zinf$ derived from the four-band model for pristine BLG has an appreciable electron density dependence. In particular, we find that
\begin{align}
	\zinf^\text{(BLG-4band)} = \dfrac{1}{2}(5\eta^2-2\eta+1). \label{Eq:zinf-BLG-4band}
\end{align}

\subsubsection{Two-band model for TDBG}
Finally, we consider a low-energy two-band description of TDBG.
Because we are only interested in the chiral form factor, we do not need the exact two-band model for TDBG.
Instead, we know from symmetry considerations that to leading order the two-band model for TDBG must be of the form
\begin{align}
	H_\text{TDBG} = \mathcal{A} \TwoDMatrix{0}{(k_x+ik_y)^2}{(k_x-ik_y)^2}{0}, \label{Eq:t-DBLG}
\end{align}
where the coefficient $\mathcal{A}$ depends on band structure details, which we do not need.
However, we do need explicit expressions for the basis states of this two-band Hamiltonian at $k=0$, which were already given in Eq.~\eqref{Eq:ZeroEnergyState} as $\ket{\Psi^{(A)}}$ and $\ket{\Psi^{(B)}}$.
With this knowledge, we can write down general expressions for the eigenstates of this two-band model at small $\bk$ as follows,
\begin{align}
	\ket{\zeta, \bk} &= \dfrac{1}{\sqrt{2}}\left(\ket{\Psi^{(A)}} + \zeta e^{-2i\phik}\ket{\Psi^{(B)}} \right) \notag\\
	&\equiv
	\begin{pmatrix}
		\ket{\zeta, \bk}_0, & \ket{\zeta, \bk}_1, & \ket{\zeta, \bk}_2, &\ket{\zeta, \bk}_3
	\end{pmatrix}, \label{Eq:TwobandEigenstate}
\end{align}
where $\zeta = \pm1$ is the band index, and
\begin{align}
	\ket{\zeta, \bk}_n = \dfrac{1}{\sqrt{2}}\left(S_A\ket{\psi_n^{(A)}} + S_B\zeta e^{-2i\phik}\ket{\psi_n^{(B)}} \right),
\end{align}
where $n = 0, 1, 2, 3$.
One can verify that $\ket{\zeta = \pm1, \bk}$ are indeed the two eigenstates of the two-band Hamiltonian Eq.~\eqref{Eq:t-DBLG}.

When calculating the chiral form factor, we will consider phonon scattering in the two layers independently.
In particular, note that the first component of the four-component eigenstate $\ket{\zeta = \pm1, \bk}$ in Eq.~\eqref{Eq:TwobandEigenstate} resides in the top BLG, while the other three reside in the bottom one.
As a result, the chiral form factor should be evaluated as follows,
\begin{align}
	F(\phi) &= \lvert _0\inner{\zeta', \bk'}{\zeta, \bk}_0\rvert^2  + \bigg\lvert \sum_{j=1}^{3} {}_j\inner{\zeta', \bk'}{\zeta, \bk}_j \bigg\rvert^2 \notag\\
	&\equiv |\mathcal{F}_1(\phi)|^2 + |\mathcal{F}_2(\phi)|^2.
\end{align}
The above derivations lead to the following results,
\begin{align}
	\mathcal{F}_1(\phi) = \dfrac{1}{2}\left(1+\dfrac{\zeta\zeta'}{1+3\Delta}e^{2i\phi}\right), \;
	\mathcal{F}_2(\phi) = \dfrac{\zeta\zeta'}{2}\dfrac{3\Delta}{1+3\Delta}e^{2i\phi},
\end{align}
which then gives rise to the following form factor for the two-band model of TDBG,
\begin{align}
	F(\phi) = \dfrac{1}{4}\left[1+\dfrac{9\Delta^2+1}{(3\Delta+1)^2} + \dfrac{2}{3\Delta+1}\cos(2\phi) \right]. \label{Eq:TDBG-FormFactor}
\end{align}
Such a chiral form factor yields the following result for $\zinf$,
\begin{align}
	\zinf^\text{(TDBG)} = 1 + \dfrac{9\Delta^2+1}{(3\Delta + 1)^2}. \label{Eq:zinf-t-dBLG}
\end{align}

\begin{figure}[!]
\includegraphics[width = 8cm]{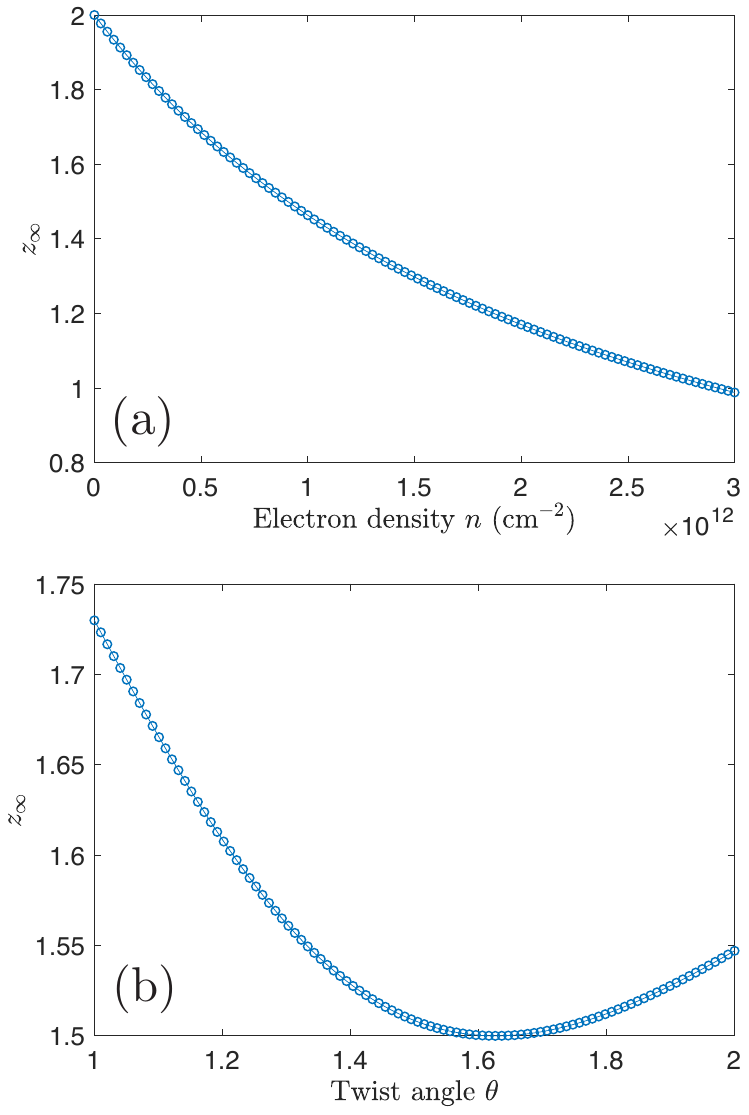}
\caption{\label{Fig:FormFactor}
Plot of $\zinf$ in (a) the four-band model for pristine bilayer graphene (i.e., $\zinf^\text{(BLG-4band)}$) and (b) the two-band model for TDBG (i.e., $\zinf^\text{(TDBG)}$). The set of parameters used in this figure is the same as that in Fig.~\ref{Fig:Bandstructure}. As a comparison, note that within the two-band model of pristine bilayer graphene, we have $\zinf^\text{(BLG-2band)} = 2$.
}
\end{figure}

Some numerical results for $\zinf$ under different approximations are given in Fig.~\ref{Fig:FormFactor}.
One can see that the three different approximations of $\zinf$ are of the same order, although they capture different aspects of the band structure.
In the rest of the paper, we will use both $\zinf^\text{(TDBG)}$ and  $\zinf^\text{(BLG-4band)}$ to calculate the phonon-induced resistivity.
Note that, in order to evaluate $\zinf$ for TDBG accurately, one has to resort to full numerical evaluations from the band structure.
Although such a calculation is beyond the scope of this work, we expect that the exact value of $\zinf$ is still within the same order of magnitude as the ones we used in this work.

\subsection{Phonon-induced resistivity: High temperature limit}
After explaining the calculations of $\zinf$, we are now ready to evaluate the  phonon-induced resistivity explicitly.
In this subsection we will consider the high-temperature limit only, when the resistivity is a linear function of temperature, and then present results for the full temperature range in the next subsection.
Before showing our results, however, we make a few comments on our numerical evaluation of the coefficient $C$ using Eq.~\eqref{Eq:Coefficient_C}.
First, we will use both $\zinf^\text{(TDBG)}$ and $\zinf^\text{(BLG-4band)}$ to approximate $\zinf$, and demonstrate how different approximations affect the final value of $C$.
Second, the Fermi velocity $v_F$ will be extracted directly from the full numerical band structure of TDBG, instead of just from the two-band effective model in Eq.~\eqref{Eq:t-DBLG}.
Finally, all of our calculations are limited to carrier densities below the van Hove singularities in the band structure, because our theory will break down for higher carrier densities, when the topology of the Fermi surface is different from our assumptions because of the complications arising from van Hove singularities.
Thus, our theory is explicitly limited to low carrier densities where the Fermi level stays below the van Hove singularities.

\begin{figure}[!]
\includegraphics[width = 8cm]{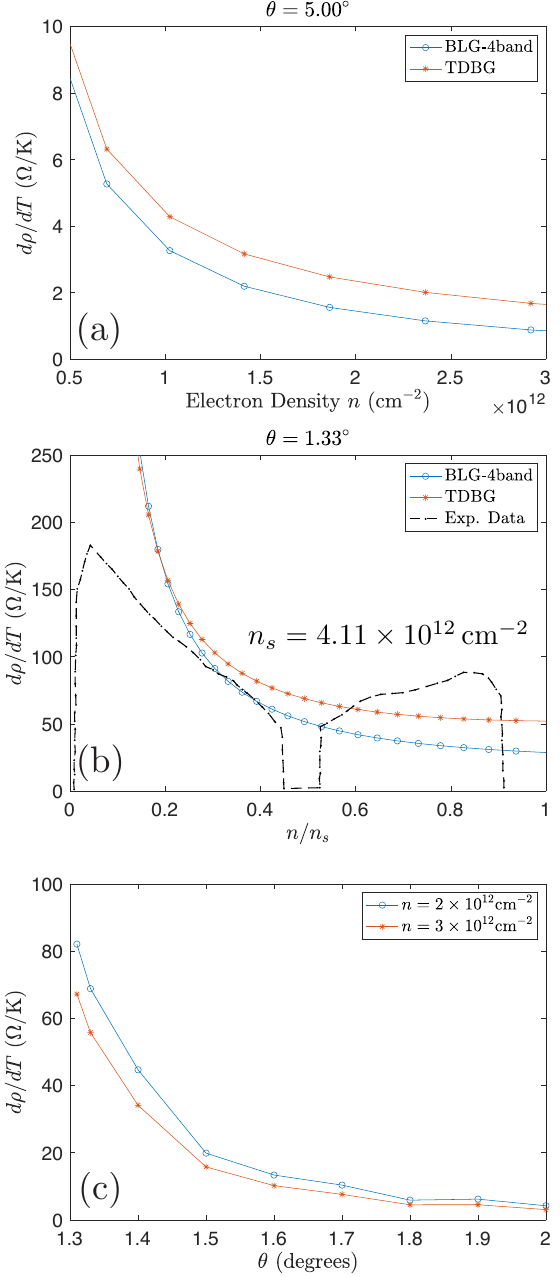}
\caption{\label{Fig:DensityDependence}
(a)-(b) Slope of the linear-in-$T$ phonon-induced resistivity for two different twist angles. The circles and stars represent the results obtained with two different approximations of $\zinf$, respectively. 
The experimental data in (b) were adapted from Ref.~\cite{Burg2019}. 
(c) Slope of the linear-in-$T$ phonon-induced resistivity as a function of twist angle at two fixed electron densities. 
Note that in this panel the results are obtained by using $\zinf^\text{(TDBG)}$ as the form factor. 
{In addition, a constant deformation potential of $D=\SI{25}{eV}$ is used for this calculation. Since the actual deformation potential in TDBG can vary between different devices (and hence between different twist angles), the quantitative behavior of $d\rho/dT$ as a function of twist angle may not be captured accurately. However, the qualitative trend that at a fixed carrier density $d\rho/dT$ increases with decreasing $\theta$ shall stand. }
}
\end{figure}

Some numerical results for the coefficient $C$ are given in Fig.~\ref{Fig:DensityDependence}.
In (a)-(b) the results for two different twist angles are shown. One can see that as the twist angle decreases from $5.0^\circ$ to $1.33^\circ$, the coefficient $C$ increases substantially.
Such a trend is also apparent in panel (c), which shows how the coefficient $C$ depends on the twist angle at a fixed electron density.
We note that such an angular dependence with resistivity increasing strongly with decreasing twist angle is consistent with recent resistivity measurements in TDBG~\cite{Shen2019,Cao2020,Liu2019}.

In addition, we find from Fig.~\ref{Fig:DensityDependence} that within our theory the coefficient $C$ has a strong density dependence, {especially at low electron densities ($n<n_s/4$, where $n_s$ is the total electron density in a moire unit cell)}.
This feature in our theory arises from the fact that the coefficient $C$ is inversely proportional to the Fermi velocity, which has a strong density dependence for parabolic bands.
{It is interesting to draw a comparison between TBG and TDBG in this context.
In particular, note that such a density dependence in $C$ is weak in the case of TBG, even within the framework of phonon-induced resistivity and at small twist angles~\cite{Wu2019}.
The underlying reason is straightforward: the low-energy electronic states in TBG has an approximate linear dispersion.
As a result, the Fermi velocity and hence the coefficient $C$ in TBG has a weak dependence on the carrier density.
It is worth noting that such a weak density dependence in $C$ is consistent with the experimental observations in TBG~\cite{Cao2018a}.
By contrast, TDBG bands are parabolic, and hence one expects a density dependence in the temperature coefficient of the resistivity.
}

It is also instructive to compare our theory with existing experimental results~\cite{Shen2019,Cao2020,Liu2019,Burg2019}.
First of all, we note that our calculated TDBG resistivity approximately agrees with recent measurements~\cite{KimPrivate} at various carrier densities with $n>n_s/4$.
For example, we find that $d\rho/dT$~$\sim$~$\SI{95}{\Omega/K}$ at a twist angle of $\theta\sim 1.24^\circ$ and a carrier density of $\SI{3.0e12}{cm^{-2}}$, compared well with the experimental values of $d\rho/dT$~$\sim$~$\SI{75}{\Omega/K}$.
{In addition, in another recent experiment~\cite{Burg2019} it was found that in a TDBG sample with $\theta = 1.33^\circ$ the value of $d\rho/dT$ varies between $\SI{50}{\Omega/K}$ and $\SI{100}{\Omega/K}$ for carrier densities $0.2<n/n_s<0.4$ and $0.6<n/n_s<0.8$, see Fig.~\ref{Fig:DensityDependence}(b). 
In comparison, this measurement is qualitatively captured by our theory. 
}

On the other hand, we find that we cannot fully capture the density dependence of $C$ observed in the experiments, although the experimental results are not unanimous at the moment either.
In particular, our theory shows that $d\rho/dT$ varies strongly with the electron density, especially at low carrier densities ($n<n_s/4$).
On the experimental side, two earlier experiments~\cite{Cao2020,Liu2019} show that $d\rho/dT$ has almost no dependence on the carrier density at low carrier densities and small twist angles ($\theta\sim 1^\circ$).
In a more recent experiment~\cite{Burg2019} {see a quotation of their results in Fig.~\ref{Fig:DensityDependence}(b)}], however, $d\rho/dT$ is shown to depend on the electron density in both TDBG samples they studied (with a twist angle of $\theta=1.01^\circ$ and $\theta=1.33^\circ$, respectively).
We speculate that one of the main reasons for such a discrepancy between our theory and the experimental results is that we did not include electron correlations in our theory, as the purpose of this study is to demonstrate that phonon scattering alone can produce a linear-in-$T$ resistivity at high temperatures along with the correct magnitude for $d\rho/dT$.
As a result, the variation of $d\rho/dT$ with the electron density may be overestimated in our theory, especially at low carrier densities ($n/n_s<0.2$).
We expect that for larger twist angles ($\theta\gtrsim 2.0^\circ$), when the band width becomes large, electron-phonon scattering will overcome the electron correlation effects, and become the dominant mechanism for resistivity in the high-temperature limit.
In that regime, we expect that $d\rho/dT$ will exhibit a much stronger carrier density dependence in the experiments.
It will thus be interesting to carry out an experiment to resolve the crossover between these two regimes, which will help us better understand the role of electron correlation in TDBG.

{
Given the lack of experimental consensus on the exact behavior of $d\rho/dT$ as a function of carrier density, it is difficult to determine the cause of the current discrepancy between theory and experiments.
Any definitive agreement between our current theory and the measured TDBG temperature dependent resistivity awaits a careful experimental study of temperature, twist angle, and density dependence of TDBG resistivity, which is unavailable right now.
}

\subsection{Phonon-induced resistivity: Full temperature range}
Finally, we evaluate the phonon-induced resistivity in the full temperature range.
Such a calculation will not only allow us to present a complete result for the phonon-induced resistivity in TDBG, but also help us determine the temperature range in which the resistivity is linear in $T$.

\begin{figure}[!]
\includegraphics[width = 8cm]{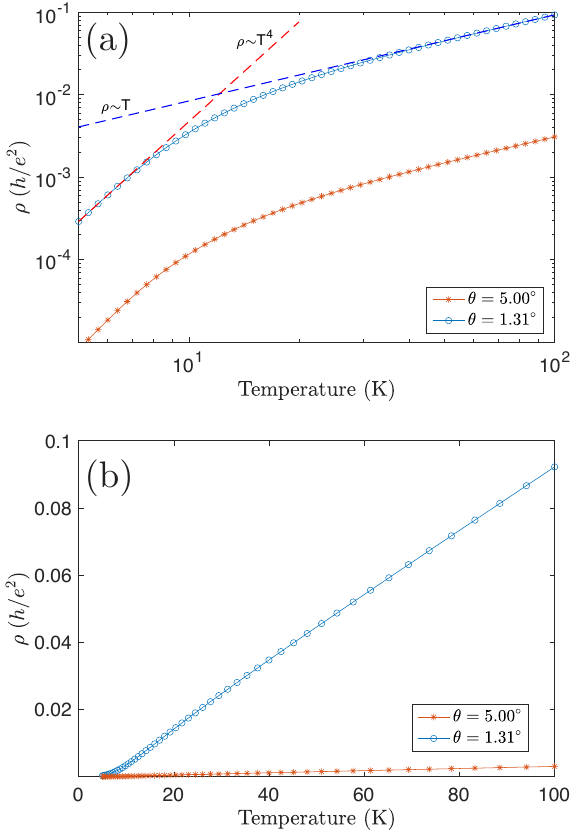}
\caption{\label{Fig:FullResistivity} (a) Phonon-induced resistivity in TDBG across the full temperature range. The results for two different twist angles are shown, and the electron density is taken to be $n=\SI{3.0e12}{cm^{-2}}$ for both cases. These results are evaluated by using the chiral form factor for the two-band model of TDBG in Eq.~\eqref{Eq:TDBG-FormFactor}. The red and blue dashed line represents a $\rho$~$\sim$~$T^4$ and $\rho$~$\sim$~$T$ asymptotic behavior, respectively.
(b) {The result in (a) is shown here in a linear scale to make it easier to make comparisons with experimental results.} }
\end{figure}

To begin with, we consider the integral $I$ in the low-temperature limit ($T\ll\TBG$), which can be evaluated by introducing $y = z_\text{BG} x$, yielding
\begin{align}
I \approx \dfrac{16}{\pi z_\text{BG}^4}	\int_0^{+\infty} \dfrac{y^4 e^{y}}{(e^y-1)^2} dy = \dfrac{16}{\pi} \times \dfrac{4! \zeta(4)}{z_\text{BG}^4} \propto T^4,
\end{align}
where we have used $F(0)\approx 1$, and $\zeta(s)$ is the Riemann-$\zeta$ function.
As a result, in the low-temperature limit the resistivity depends on the temperature as $\rho \propto T^4$.
This is the so-called {\BGBGBG}   regime of phonon scattering, which in 3D metals produces a $T^5$ power law for the temperature-dependent resistivity.
In addition, the resistivity becomes independent of the chiral form factor $F(\theta)$ in this low-temperature limit.
In Fig.~\ref{Fig:FullResistivity} we show the resistivity in TDBG across the full temperature range for a carrier density of $\SI{3.0e12}{cm^{-2}}$ and two different twist angles.
These results are obtained by using the chiral form factor for the two-band model of TDBG in Eq.~\eqref{Eq:TDBG-FormFactor}.
One can clearly observe that the resistivity has a $\rho \propto T^4$ behavior in the low-temperature range, while a $\rho\propto T$ behavior in the high-temperature limit.

\begin{figure}[!]
\includegraphics[width = 8cm]{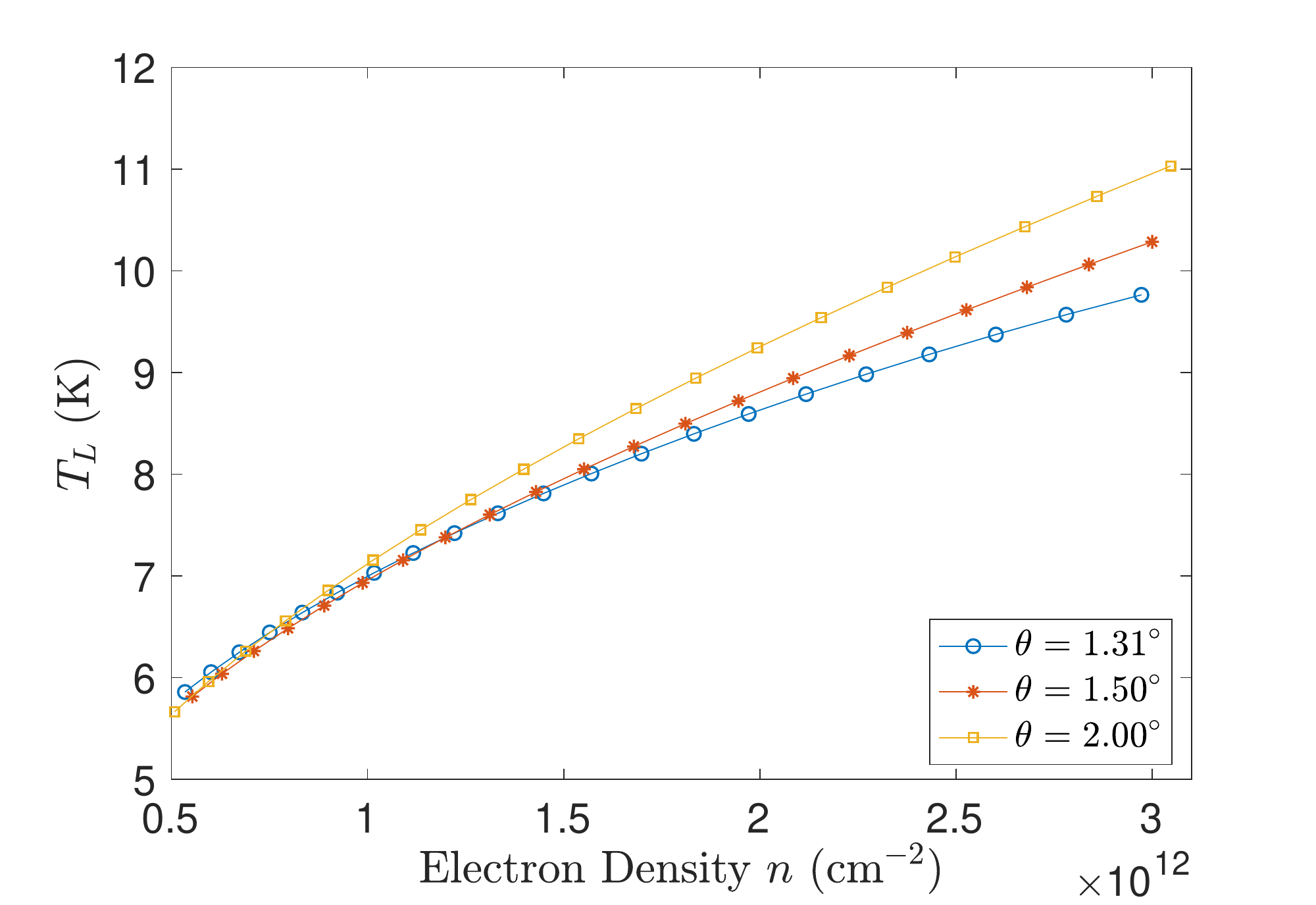}
\caption{\label{Fig:TL}
Critical temperature $T_L=\TBG/4$ as a function of electron density for three different twist angles.
}
\end{figure}

After obtaining the resistivity in the full temperature range, it is instructive to examine the crossover temperature above which the resistivity becomes linear in temperature.
It has been established previously that the linear-in-$T$ behavior already kicks in at a characteristic temperature $T\gtrsim T_L\approx\TBG/4$~\cite{Hwang2008,Min2011}.
In Fig.~\ref{Fig:TL} we plot the crossover temperature $T_L$ in TDBG as a function of carrier density for three different twist angles. We find that $T_L$ is below $\SI{11}{K}$ for almost all carrier densities and twist angles we considered.
As a result, our analysis of the phonon-induced resistivity in the high temperature limit should be applicable above $\sim \SI{11}{K}$.
In fact, recent resistivity measurements in TDBG indeed show  a linear-in-$T$ behavior for temperatures between $\SI{10}{K}$ and $\SI{30}{K}$~\cite{KimPrivate}, a temperature range where our theory is applicable.
Therefore, our theory will be relevant for the understanding of the linear-in-$T$ resistivity observed in recent experiments in TDBG~\cite{Shen2019,Cao2020,Liu2019}.
We note that in the low-temperature {\BGBGBG}  regime, the $T^4$ power law in the resistivity may not be easy to discern because of other resistive scattering contributions such as electron-impurity and electron-electron interactions which are neglected in our theory.

\section{Phonon-Mediated Superconductivity\label{Section:Discussions}}

We now discuss possible implication of our theory on superconductivity in TDBG. The electron-acoustic phonon coupling mediates an effective attractive electron-electron interaction with a strength given by $g_0$~$=$~$D^2/(4 \rho_m \vph^2) \approx \SI{50}{meV\cdot nm^2}$~\cite{Wu2019}.
The dimensionless electron-phonon coupling constant is determined by $\lambda^* = g_0 \nu(\varepsilon_F)$, where $\nu(\varepsilon_F)$ is the DOS per spin and valley.
Because of the narrow bandwidth for small twist angle ($\sim 1^{\circ}$), $\lambda^*$ in TDBG can reach order of $0.25$ given the DOS shown in Fig.~\ref{Fig:Bandstructure}(d).
The superconducting transition temperature $T_c$ can be roughly estimated as $k_B T_c = \Lambda \exp(-1/\lambda^*)$ within a BCS-type theory, where $\Lambda$ is a cutoff energy approximately given by the flatband bandwidth ($\sim 5$ meV).
Therefore, $T_c$ can be of order $\SI{1}{K}$ from electron-phonon interactions. Moreover, the effective attractive interactions mediated by acoustic phonons have an enlarged SU(2)~$\times$~SU(2) symmetry, namely, each valley has its own spin rotational symmetry. Therefore, acoustic phonons mediate both spin singlet and spin triplet pairings~\cite{Wu2019}, and can account for the spin triplet superconductivity  experimentally identified in TDBG~\cite{Liu2019}.
We note that the possibility of the electron-phonon mediated superconductivity at low temperatures ($<\SI{1}{K}$) and the large phonon-induced linear-in-$T$ resistivity at high temperatures ($>\SI{10}{K}$) are closely connected, both arising from the strongly enhanced electron-phonon coupling induced by flatband moire physics, as has already been emphasized in the context of TBG in Ref.~\cite{Wu2019}.

\section{Discussions and Conclusions \label{Section:Conclusions}}

{Before we conclude, we explain why the level of phonon model we adopt is adequate for our purpose. 
We start by noting that in our theory only two quantities depend on the phonon model: the sound velocity and the electron-phonon coupling. 
As we now explain, the sound velocity is not substantially affected by the {\Moire} potential, while the electron-phonon coupling should be regarded as an unknown fitting parameter, as it is almost impossible to determine without experimental inputs. 

To begin with, we explain why the sound velocity is in principle not substantially affected by the {\Moire} potential. 
We believe that the acoustic phonons are quite similar in TDBG and TBG, and the
latter has been analyzed in the literature. 
For example, the low-energy phonon spectrum (in particular the sound velocity) in TBG has been analyzed in Ref.~\cite{Koshino2019}. 
It was shown that the {\Moire} interlayer potential gives rise to
superlattice zone folding, which resulted in the appearance of minibands in the
phonon spectrum. 
In particular, in the long wavelength limit (the limit we are
working in) the phonon spectrum in TBG only consists of two branches, just like
what happens in two decoupled graphene sheets. Furthermore, it was found that as
the twist angle is reduced to $1^\circ$, the two sound velocities are
only reduced by 20\%-30\% compared to their values in a pristine single-layer
graphene. 
Thus, the impact of the {\Moire} structure on the sound velocities
is generally not important.

Next, we note that it is almost impossible to estimate the electron-phonon coupling without experimental inputs. 
On the one hand, there is no consensus on the precise value of the deformation potential even in pristine graphene (see the new references we added). The commonly quoted values vary between \SI{10}{eV}-\SI{40}{eV}~\cite{Pietronero1980,Sugihara1983,Woods2000, Suzuura2002, Pennington2003,Stauber2007, Vasko2007}, which already resulted in a substantial uncertainty. 
On the other hand, this coupling can be renormalized by effects like electron-electron interactions, which is not taken into account in our theory to begin with. 
The above discussion not only applies to our system, but also to most electronic materials (e.g. regular monolayer graphene and bilayer graphene) because estimating the deformation potential in real samples accurately in theory is essentially an impossible task. 
In fact, resistivity measurements are often used to estimate the electron-phonon coupling strength in metals and semiconductors. 
Consequently, we believe that there is no point in trying to calculate the deformation potential accurately. Instead, we leave it as an unknown fitting parameter, which should be extracted from experimental results. 

Since our electron-phonon coupling strength is a tuning parameter to be obtained by comparing with experiments (and right now there are experimental variations of unknown origin since the resistivity measured in different laboratories do not quite agree), 
it makes little sense to discuss whether we are considering deformation potential coupling or the so-called ``gauge coupling'' (associated with the phonon-induced variations in the off-diagonal terms in the electron hopping parameter~\cite{Park2014,Pietronero1980,Sugihara1983,Woods2000}, 
since the transport theory for the two cases are identical~\cite{Hwang2008,Park2014} with the only difference being that the electron-phonon coupling strength is calculated using different microscopic theories in the two cases (and we are not calculating the electron-phonon coupling microscopically, we are choosing it as a phenomenological parameter). 
We argue below that screening plays no role in determining the electron-phonon coupling strength, and therefore a single parameter $D$ can represent the overall electron-phonon coupling strength in our theory. 
Future improvements of our theory may take into account the microscopic aspects of electron-phonon coupling in real TBG or TDBG, but such theories will be highly numerical and must take into account the random strain in the twisted structures, which may turn out to be a huge challenge in calculating the coupling strength from first principles.

It is also instructive to discuss the applicability of our theory. 
First of all, electron-electron interaction is not accounted for in our theory. 
As a result, when the conduction band is nearly empty, half-filled, or full, our theory is likely to fail, as these are the density ranges in which correlation effects are strong, as can be seen from the experimental results quoted in Fig.~\ref{Fig:DensityDependence}(b). 
Moreover, our theory does not capture the physics near Van Hove singularities in the band structure, such as the ones shown in the density of states plots in Fig.~\ref{Fig:Bandstructure}(c)-(d), as our theory is based on a parabolic band approximation. 
However, it is difficult to give a quantitative estimate on the exact range of carrier densities in which our theory applies, as many factors can come into play. 
In our opinion, direct comparisons with experimental results (like the one shown in Fig.~\ref{Fig:DensityDependence}) will provide the best test of our theory.

One question which is relevant here is the possible role electron-electron interactions (and consequentially, screening) play in determining the electron-phonon coupling strength in TDBG (or for that matter, twisted bilayer graphene, monolayer graphene, and bilayer graphene). 
This is a question of some importance since electron-electron interactions are strongly enhanced in twisted {\Moire} systems because of the suppression of the Fermi velocity. 
A complete answer to this question, which has recently been discussed also in Ref.~\cite{Sarma2020}, is well beyond the scope of the current work, and has not been attempted in any theoretical work in the graphene literature. 
In fact, the microscopic interplay of electron-electron interaction, screening, and electron-phonon interaction is extremely difficult even in simple metals where the effect is often subsumed in an unknown phenomenological parameter, called $\mu^\ast$, in the context of theories on superconductivity. 
Obviously, this is not a problem, which can be solved theoretically at any level of rigor. 
Our approach in the current work, following the highly successful similar theories in Refs.~\cite{Hwang2008,Min2011,Wu2018,Wu2019} is to take the basic electron-phonon coupling strength as an unknown parameter to be determined from experiments, where this coupling strength (``deformation potential'') sets the overall scale of the high-temperature linear-in-$T$ resistivity. 
This approach has not only been highly successful in graphene, but also has been used extensively in semiconductors and metals. 
We therefore believe that such a phenomenological approach is the appropriate theoretical approach at this early stage of the subject. 
There are very good reasons to believe that screening plays no role in the theory developed in this paper, and we elucidate these reasons in the next two paragraphs.

One reason for the irrelevance of screening is provided by recent experiments~\cite{Stepanov2019,Saito2020,Liu2020}  which directly studied the screening effects in twisted graphene systems by tuning nearby screening gates. 
The experimental conclusion of these direct twisted graphene screening experiments is unambiguous---it shows that neither superconductivity nor the linear-in-$T$ high-temperature resistivity are affected by screening in contrast to the correlated insulator phase which disappears under strong screening. 
One therefore must conclude that screening does not affect the scattering mechanism controlling the high-temperature resistivity which is the main subject matter of our work. 
Thus, phonon scattering is unaffected by screening according to direct experimental investigations. 
This provides strong empirical support for our phenomenological electron-phonon interaction model ignoring any screening by electron-electron interactions.

The second reason for our neglect of screening is that the electron Fermi velocity in the flatband twisted {\Moire} graphene system is smaller than the phonon velocity in sharp contrast to ordinary metals where the Fermi velocity is orders of magnitude larger than the phonon velocity. 
In such a situation, instead of static screening, the screening by electrons should be highly dynamical since the phonons are moving much faster than the electrons (i.e. the opposite of what happens in metals). 
The dynamical screening situation usually implies anti-screening rather than screening, i.e., the dynamically screened  interaction is in fact stronger than the unscreened interaction. 
Hence, ``screening'' in our system may even lead to an enhanced effective electron-phonon interaction instead of a weakened electron-phonon interactions. 
It is therefore possible for the effect studied in our work to be enhanced by electron-electron interactions, leading to an effectively larger deformation potential coupling. 
This issue should be revisited in future theories once the experimental resistivity measured in different laboratories agrees with each other necessitating a quantitatively accurate theoretical description.

We also emphasize that there exists important differences between the phonon scattering induced resistivity in TBG and TDBG. 
In particular, our work shows that $d\rho/dT$ in TDBG has a strong dependence on the carrier density, which is completely absent in the corresponding theory for TBG. 
This qualitative difference is one of the key findings in our work, and it provides an important theoretical reference for ongoing experiments. 
Future experiments should study the density dependent resistivity carefully in order to verify our predictions.
Moreover, in contrast to TBG, the TDBG studied in this work is characterized by an effective mass rather than a constant Fermi velocity.  
In addition, {TDBG does not have a magic angle whereas TBG does}. 
These differences make the current TDBG theory completely different from earlier theories on phonon scattering in TBG. 
}

To summarize, in this work we developed a theory to calculate the phonon-induced resistivity in twisted double bilayer graphene in the high-temperature ($T>\TBG$) limit, where the resistivity $\rho$ scales linearly with temperature $T$, $\rho \approx CT$.
We present a quantitative analysis of the coefficient $C$ and showed that it increases substantially as the twist angle $\theta$ is reduced.
However, since we did not account for electron correlation effects on the resistivity, we expect that our predictions are likely applicable for devices in which the twist angle is relatively large ($\theta\gtrsim 2^\circ$).
The main qualitative conclusion of our theory is that for $T>\SI{10}{K}$ or so, TDBG should manifest a very large linear-in-$T$ resistivity arising from phonon scattering at small twist angles.
The linear coefficient should manifest a strong density dependence, which is not seen in current experiments for reasons not obvious right now.

\section*{Acknowledgment}
This work is supported by Microsoft and Laboratory of Physical Sciences.
X.L. also acknowledges support from City University of Hong Kong (Project No. 9610428).

\bibliographystyle{apsrev4-1}
\bibliography{Ref-TDBLG}

\end{document}